\documentclass[aps,pra,twocolumn,showpacs,superscriptaddress]{revtex4}
\usepackage{epsfig,graphics}
\usepackage{float}
\usepackage{bm,amsmath,amssymb,latexsym,graphicx,enumerate}
\usepackage[mathcal]{euscript}
\usepackage{hyperref}

\newcommand{\be}{\begin{equation}}
\newcommand{\ee}{\end{equation}}

\begin{document}

\title{Polarization rotation by an rf-SQUID metasurface}

\author{J.-G.~Caputo$^{~1}$, I. Gabitov$^{~2}$  and A.I.~Maimistov$^{~3,4}$}

\affiliation{\normalsize \noindent
$^1$: Laboratoire de Math\'ematiques, INSA de Rouen, \\
Avenue de l'Universite,
Saint-Etienne du Rouvray, 76801 France \\
E-mail: caputo@insa-rouen.fr \\
$^2$: Department of Mathematics,\\
University of Arizona, Tucson, AZ, 85704, USA \\
E-mail: gabitov@math.arizona.edu\\
$^3$: Department of Solid State Physics and Nanostructures, \\
National Research Nuclear University, Moscow Engineering Physics Institute, \\
Kashirskoe sh. 31, Moscow, 115409 Russia \\
E-mail: amaimistov@gmail.com \\
$^4$: Department of Physics and Technology of Nanostructures \\
Moscow Institute for Physics and Technology, \\
Institutskii lane 9, Dolgoprudny,
Moscow region, 141700 Russia \\}

\date{\today }

\begin{abstract}
We study the transmission and reflection of a plane electromagnetic wave
through a two dimensional array of rf-SQUIDs. The basic equations
describing the amplitudes of the magnetic field and current in the
split-ring resonators are developed. These yield in the linear approximation
the reflection and transmission coefficients. 
The polarization of the reflected wave is independent of the
frequency of the incident wave and of its polarization; it is defined
only by the orientation of the split-ring. The reflection and
transmission coefficients have a strong resonance that is determined by
the parameters of the rf-SQUID; its strength depends essentially on
the incident angle. 
\end{abstract}

\pacs{Josephson devices, 85.25.Cp, Metamaterials  81.05.Xj,
Microwave radiation receivers and detectors, 07.57.Kp}
\maketitle

\section{Introduction }

Recently a layer of metamaterial containing specially etched designs,
referred to as metasurfaces, was used to induce a phase gradient in
an incident electromagnetic wave\cite{ygk11}. This leads to a
generalized Snell's law and the control of the transverse
structure of the wave front\cite{ygk11}. This study was
done in the optical domain, it then was extended to microwaves
by Shalaev et al \cite{shalaev12}. The phase gradient is due
to a plasmon resonance between the electromagnetic wave and the
designs etched on the surface. These have to be adapted for
each frequency domain and are fixed by construction.

It would be
useful to have a system whose response could be changed over a
significant range of frequencies. Such a
device exists, it is a split-ring Josephson resonator (rf-SQUID)
\cite{Lazarides:07a,Lazarides:07b,Lazarides:09,Maim:Gabi:10}.
The rf-SQUIDs as basic elements of quantum metamaterials were
discussed in \cite{Chunguang:06,Anlage:11,Trepanier:11,cgm12}. In
the experimental study \cite{Ustinov:13}, it was shown that one
can tune the resonance frequency of these devices.
It should be pointed that this system has a discrete energy spectrum
and a large magnetic momentum \cite{cgm12}. Then the energy of
interaction with an external field can be of the order of the
transition energy between neighboring energy states.

In this article we show that a film of properly oriented rf-SQUID
controls the polarization of a wave reflecting on the
meta-surface. This is similar to a Faraday effect; the wave
is strongly reoriented at the resonance.
We determine the parameters of the reflected and transmitted
waves in the linear approximation. The reflection and transmission coefficients
depend on the frequency and on the stationary state of the system.
In that sense, the device is active, its parameters can be modified.

\section{The model}

\begin{figure}
\centerline{ \resizebox{8 cm}{5 cm} {\includegraphics{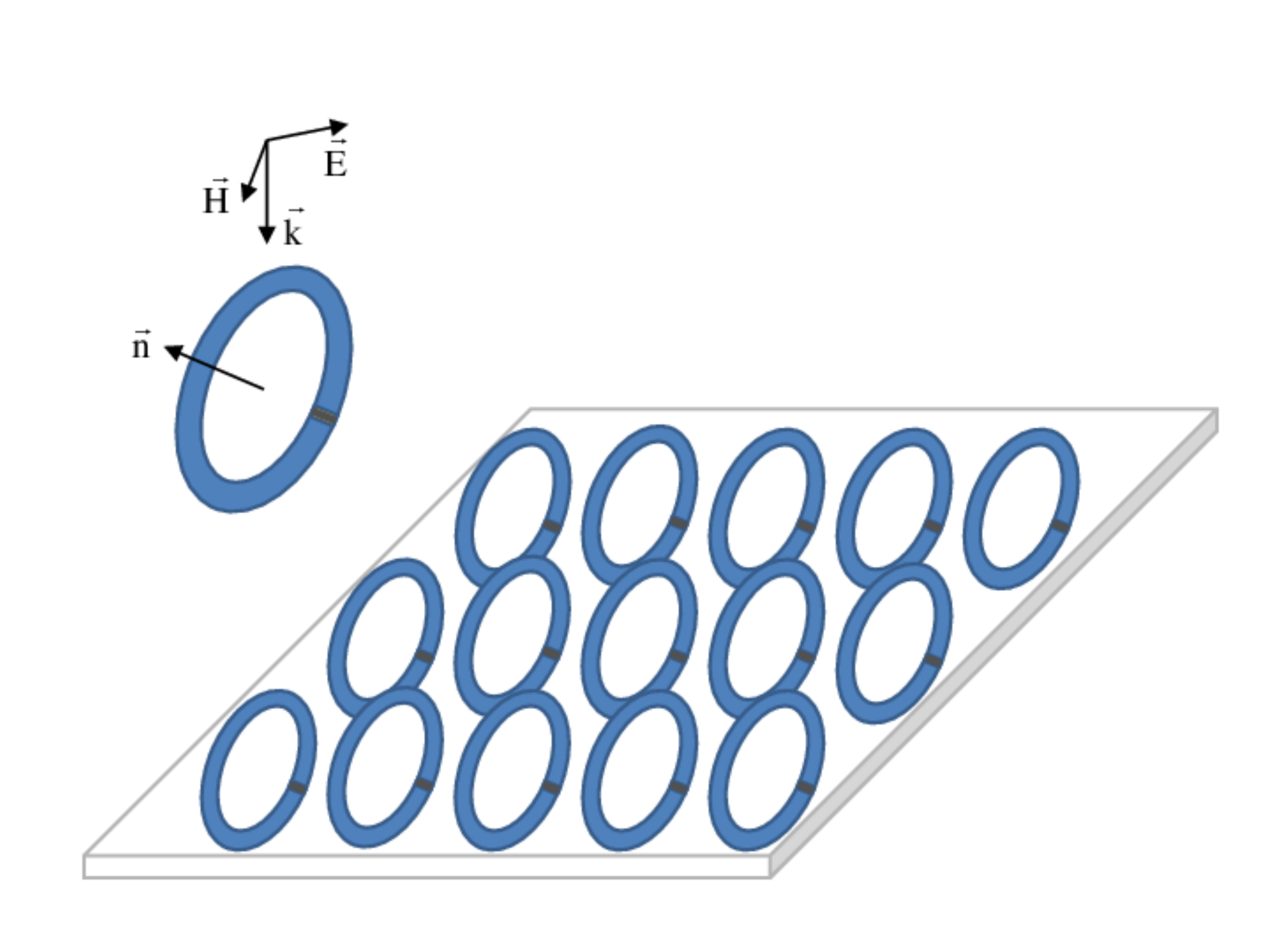} }}
\caption{
Schematic view of the electromagnetic wave
incident on the layer containing split ring Josephson junction
resonators. One ring is shown, indicating the orientation.}
\label{f1}
\end{figure}

We consider a plane wave normally incident on a
layer of rf-SQUIDs as shown in Fig.\ref{f1}. All the rings in the
layer are oriented in the same direction given by the normal
vector $\mathbf{n}$. The interaction of the
individual rf-SQUID with the electromagnetic field is determined by the
magnetic flux through the split ring. Therefore the orientation of
the rf-SQUID controls the parameters of the transmitted and reflected waves.
This orientation is characterized by the angle $\theta$ between
the magnetic component $\mathbf{H}$ and the normal $\mathbf{n}$ to the split
ring. We assume the same homogeneous dielectric layers above and below the
the film.
The dielectric
permittivity  of this medium is $\varepsilon$.  The model describing
the interaction of electromagnetic field with the system of rf-SQUIDs is
based on Maxwell's equations and the equation for the response of the
rf-SQUIDs:
\begin{eqnarray} 
&& \nabla\times \mathbf{E} = -(\mu_0 \mathbf{H}+ \mathbf{M})_t, \label{m1}\\
&& \nabla\times \mathbf{H} = \varepsilon_0 \varepsilon \mathbf{E}_t, \label{m2} \\
&& \nabla\cdot \mathbf{D} =0, \quad \nabla\cdot \mathbf{B} =0. \label{m3}
\end{eqnarray}
The magnetization $\mathbf{M}$ in the equation~(\ref{m1}) is
localized in the array whose thickness is much smaller than
the wavelength $\lambda$. Therefore the magnetization can be
written as follows:
\be\label{magnetization}
\mathbf{M}(t,\mathbf{r})=\sum_{a}\mathbf{m}_{a}(t)\delta(z) \approx \mathbf{m}(t)n_r l\delta(z),
\ee
where $\mathbf{m}_{a}(t)$ are individual rf-SQUIDs  magnetizations, $l<<\lambda$\ is the film thickness, $n_r$\ is the density of rf-SQUIDS
resonators and $\mathbf{m}(t)$ is the magnetic moment of
the ring \cite{Maim:Gabi:10}.
From Maxwell's equations~(\ref{m1}-\ref{m3}) we obtain the wave equation
\be
\label{gen_wave2}
\mathbf{H}_{tt}
- {c^2 \over \varepsilon} \Delta \mathbf{H} =
-{1 \over \mu_{0} } \mathbf{M}_{tt}.
\ee
Since the model considered is
translationally invariant with respect to the $x$ axis,
this equation  can be presented as
\be\label{wave3}
\mathbf{H}_{tt}
- {c^2 \over \varepsilon} \mathbf{H}_{zz} =
-{1 \over \mu_{0} } \mathbf{M}_{tt}.
\ee
Let us now consider the Josephson split-ring part. The magnetic moment is
\be\label{mag_moment}
\mathbf{m}(t) = S I \mathbf{n} ,\ee
where $I$ is the current in the loop (\ref{current}).
Combining the two equations (\ref{magnetization}) and
(\ref{mag_moment}) we get the
final expression for $\mathbf{M}$
\be\label{magnetization2} \mathbf{M} =  S I n_r l\delta(z)
\mathbf{n}.
\ee
Plugging the above expression into the wave equation (\ref{wave3})
yields the wave equation
\be\label{wave4}
\mathbf{H}_{tt}
- {c^2 \over \varepsilon} \mathbf{H}_{zz} =
-{1 \over \mu_{0} } S I_{tt} n_r l\delta(z) \mathbf{n}.  \ee

As in \cite{cgm12} the current in the ring is given by
\begin{equation} \label{current1}
I=-L^{-1}(\Phi+2 \pi \Phi_{0}\varphi),
\end{equation}
where $\Phi$ is the flux induced by the electromagnetic field,
where $L$ is the inductance of the loop, where
$\varphi$ is the supraconducting phase in the junction and
where $\Phi_{0}={\hbar \over 2 \pi}$ is the reduced flux quantum.
The flux $\Phi$ across the ring of area $S$ is
$$\Phi = S \mathbf{H} \cdot \mathbf{n} $$
where $\mathbf{n}$ is the normal vector to the ring.
This gives us the current in the ring
\be \label{current}
I= -L^{-1} \left ( S \mathbf{H} \cdot \mathbf{n}
+ 2 \pi \Phi_{0}\varphi \right ) .\ee
The evolution of the variable $\varphi$ is the same as
in \cite{cgm12} except that the current on the right hand side is modified.
We have
\be\label{phit}
2 \pi C\Phi_{0}\frac{\partial^{2}\varphi}{\partial t^{2}}
+ 2 \pi {\Phi_{0}\over R}\frac{\partial\varphi}{\partial t}
+ I_c\sin\varphi
= I.\ee
Plugging (\ref{current}) into (\ref{phit}) and multiplying by
$L /(2 \pi \Phi_{0})$ we get
\be\label{jj2}
L C \varphi_{tt} + {L \over R} \varphi_{t} +
{L I_c \over 2 \pi \Phi_{0}} \sin\varphi
= - {S \over 2 \pi \Phi_{0}} \mathbf{H} \cdot \mathbf{n} -\varphi . \ee

Our model consists in the wave equation (\ref{wave4}) together
with the split-ring Josephson equation (\ref{jj2}).
These can be normalized as in \cite{cgm12}.
The natural units of time, flux and space are
$$ \omega_T = 1/\sqrt{LC},~~  \Phi_0,
~~z_0 = {c \over \omega_T \sqrt{\varepsilon} },$$
where $ \omega_T $ is the
Thompson frequency and $z_0$ is the inverse of the Thompson wave number.
The magnetic field $\mathbf{H}$ is normalized as
\be\label{norm}
\mathbf{h} = { \mathbf{H} S  \over 2 \pi \Phi_{0}},~~\qquad\tau=\omega_{T}t,
\qquad\zeta=z/z_{0}.
\ee
In terms of these variables the final equations are
\be\label{wave}
\mathbf{h}_{\tau \tau}
- \mathbf{h}_{\zeta\zeta} =
\kappa \left (\mathbf{h}_{\tau \tau} \cdot \mathbf{n} +
\varphi_{\tau \tau} \right ) \delta(\zeta) \mathbf{n}.  \ee
\begin{equation}\label{jj}
\varphi_{\tau\tau}+\alpha \varphi_{\tau}+ \varphi+\beta\sin\varphi
=-\mathbf{h} \cdot \mathbf{n} ,
\end{equation}
where the parameters $\alpha,~~\beta$ and $\kappa$ are
\be\label{alfabeta} \alpha = {1\over R} \sqrt{L \over C},~~
\beta = {L I_c \over 2 \pi \Phi_{0} },~~
\kappa = { n_{r}l S^2  \over \mu_0 L z_0}. \ee

The equations (\ref{wave},\ref{jj}) are our principal model and we
now proceed to analyze them.

\section{Microwave spectroscopy data}

\noindent As in \cite{cgm12} we assume that the film is submitted to
a stationary field that will fix the phase $\varphi_s$. We can then
write
$$\mathbf{h} = \mathbf{h}_s + \delta \mathbf{h},~~\varphi = \varphi_s + \delta \varphi $$
where the field $\delta \mathbf{h}$ and the variable $\delta \varphi $
are small.  Their evolution is given by
\be\label{dwave}
\delta \mathbf{h}_{\tau \tau}
- \delta \mathbf{h}_{\zeta\zeta} =
\kappa \left (\delta \mathbf{h}_{\tau \tau} \cdot \mathbf{n} +
\delta \varphi_{\tau \tau} \right ) \delta(\zeta) \mathbf{n}.  \ee
\begin{equation}\label{djj}
\delta \varphi_{,\tau\tau}+\alpha \delta \varphi_{,\tau}+ \delta \varphi
+\beta\cos\varphi_s \delta \varphi
=-\delta \mathbf{h} \cdot \mathbf{n} .
\end{equation}

To solve the equations (\ref{dwave},\ref{djj}) we assume
the usual harmonic dependence
$$\delta \mathbf{h}  = e^{i \omega t} \mathbf{f} ,~~
\delta \varphi  = e^{i \omega t}\phi .$$ This yields the following
equations \be\label{swave} \mathbf{f}_{\zeta\zeta} + \omega^2
\mathbf{f} = \omega^2 \kappa \left (\mathbf{f} \cdot \mathbf{n} +
\phi \right ) \delta(\zeta) \mathbf{n},\ee \be\label{sjj} \phi \left
( \omega_r^2 -\omega^2   +\alpha i \omega \right ) = - \mathbf{f}
\cdot \mathbf{n} , \ee where we have introduced the resonant
frequency $\omega_r$ \be\label{reson} \omega_r^2 =1 + \beta \cos
\varphi_s  .\ee

We now set up the scattering experiment by assuming an incident field
and calculating the reflected and transmitted fields. For $\zeta<0 $ we have
\be\label{scat_zm}
\mathbf{f_-}(\zeta)  = \mathbf{f}_{\rm in}e^{-i \omega \zeta}
+ \mathbf{f}_{\rm r} e^{i \omega \zeta} . \ee
For $\zeta>0$ the field is
\be\label{scat_zp}
\mathbf{f_+}(\zeta) = \mathbf{f}_{\rm tr}  e^{i \omega \zeta} . \ee
At the interface $\zeta=0$ $\mathbf{f}$ is continuous so  that \be
\label{cont} \mathbf{f}_{\rm in}+\mathbf{f}_{\rm r} =\mathbf{f}_{\rm
tr}. \ee We have the following jump condition for $\mathbf{f}_\zeta$
\be\label{jump} \left[  \mathbf{f}_\zeta \right]_{0_-}^{0_+} =
\kappa \omega^2 ( \mathbf{f}_{\rm tr} \cdot  \mathbf{n} + \phi )
\mathbf{n}. \ee This yields \be \label{jump2} \mathbf{f_{in}} -
\mathbf{f_{r}} =\mathbf{f}_{\rm tr}-i\omega \kappa( \mathbf{f}_{\rm
tr} \cdot  \mathbf{n} + \phi ) \mathbf{n}. \ee From the
equations~(\ref{cont},~\ref{jump2}) it follows
\begin{eqnarray}
&& \mathbf{f_{in}}=\mathbf{f}_{\rm tr}-i\frac{\omega \kappa}{2}\mathcal{M}(\omega)]( \mathbf{f}_{\rm tr} \cdot  \mathbf{n}  ) \mathbf{n},\\
&& \mathbf{f_{r}}= i\frac{\omega \kappa}{2}\mathcal{M}(\omega)(
\mathbf{f}_{\rm tr}\cdot \mathbf{n}) \mathbf{n},
\end{eqnarray}
where \be\label{def_l} \mathcal{M}(\omega) =1+(\omega^2
-\omega_{r}^2 -i \alpha \omega)^{-1} . \ee From the first equation
we obtain \be \label{jump3} ( \mathbf{f}_{\rm tr} \cdot
\mathbf{n})=D(\omega)( \mathbf{f}_{\rm in} \cdot  \mathbf{n} ), \ee
where \be\label{def_d} D(\omega)=\left[1-i\frac{\omega \kappa}{2
}\mathcal{M}(\omega))\right]^{-1}. \ee Using these expressions we
obtain
\begin{eqnarray}
&&\mathbf{f_{r}}  = i \frac{\omega \kappa}{2}
\mathcal{M}(\omega)D(\omega)
(\mathbf{f}_{\rm in} \cdot \mathbf{n})\mathbf{n}, \label{fr}\\
&&\mathbf{f_{tr}} =\mathbf{f}_{\rm in}  +i \frac{\omega \kappa}{2}
\mathcal{M}(\omega)D(\omega)(\mathbf{f}_{\rm in} \cdot \mathbf{n})
\mathbf{n}. \label{ftr}
\end{eqnarray}
Equation (\ref{fr}) implies that the polarization of the reflected
wave is determined by $\mathbf{n}$ only. On the other hand the
direction of the transmitted wave depends on the orientations of
$\mathbf{n}$ and $\mathbf{f}_{\rm in}$ and on the frequency
$\omega$.

From expression (\ref{fr}) the reflection coefficient is
\begin{equation}\label{ref}
    \mathcal{R} =
\frac{|\mathbf{f_{r}}|^2}{|\mathbf{f_{in}}|^2} =
\frac{(\omega\kappa/2)^2|\mathcal{M}(\omega)|^2\cos^2\theta}{1+
(\omega\kappa/2)^2|\mathcal{M}(\omega)|^2},
\end{equation}
where $\theta$ is defined by the formula \be\label{def_theta}
\mathbf{f}_{\rm in} \cdot \mathbf{n}=|\mathbf{f}_{\rm in}|\cos\theta
. \ee The transmission coefficient is $\mathcal{T} =
|\mathbf{f_{tr}}|^2/ |\mathbf{f_{in}}|^2$. It is
\begin{eqnarray}
  \mathcal{T} &=& 1-\frac{(\omega\kappa/2)^2|\mathcal{M}(\omega)|^2\cos^2\theta}{1+
(\omega\kappa/2)^2|\mathcal{M}(\omega)|^2}\cos^2\theta+ \nonumber  \\
  ~ &~& ~~ +i\frac{\omega\kappa}{2}\frac{[\mathcal{M}(\omega)-\mathcal{M}^{*}(\omega)]}{1+
(\omega\kappa/2)^2|\mathcal{M}(\omega)|^2}\cos^2\theta = \nonumber\\
~&=& 1-\mathcal{R}
-\frac{\omega\kappa~\mathrm{Im}\mathcal{M}(\omega)}{1+
(\omega\kappa/2)^2|\mathcal{M}(\omega)|^2}\cos^2\theta
.\label{trans}
\end{eqnarray}

\begin{figure}[H]
\centerline{ \resizebox{8 cm}{5 cm} {\includegraphics{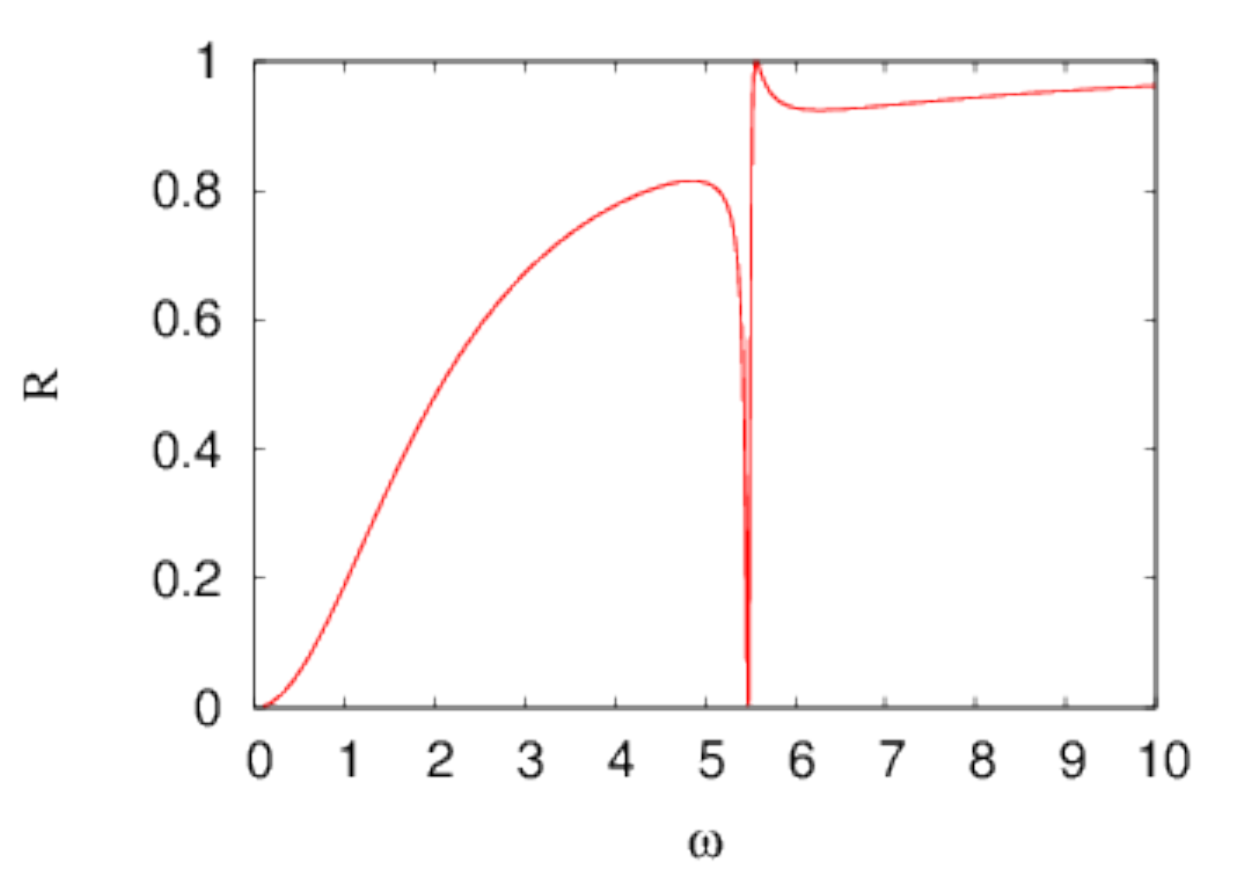} }}
\caption{The reflection coefficient $\mathcal{R}(\omega)$ from (\ref{ref}) for
a plane wave normally incident to the film,$\theta =0$.}
\label{refl}
\end{figure}
Let us analyze numerically the reflection and transmission coefficients.
We choose $\beta=30$ so that we are in the highly hysteretical
case as in \cite{cgm12}. 
Fig.  \ref{refl} shows the dependance of the reflection coefficient on $\omega$ for a normal incidence in the lossless case $\alpha=0$,  when  $\kappa=1$ 
and $\beta=30$. The external magnetic field $h_s$ is assumed to be zero. Then 
we can take $\varphi_s =0$ 
which is the global minima of the potential of the equation~(\ref{jj})(see 
Fig 2.  in~\cite{cgm12}). The reflection coefficient 
shows a strong resonance for frequencies near $\omega_r$; this resonance
is of the  Fano type
(see \cite{MFK:10}). 
The value of the resonance frequency is  
\be\label{reson2} \omega_r = \sqrt{ 1+ \beta \cos \phi_s} \approx  5.56 .
\nonumber
\ee
The metasurface is transparent for small $\omega$; when the frequency is large the incident field is totally reflected.
The shape of the spectrum does not depend on the specific value of $\beta$
and $\omega_r$.
For instance, the experimentalists \cite{Ustinov:13} chose a smaller $\beta$ and
obtain a reflection coefficient that behaves similarly to the one shown
in Fig. \ref{refl}.

The magnetic component of the reflected wave is always oriented in 
the direction of the normal. On the contrary, the polarization angle of the 
transmitted wave $\mathbf{f_{tr}}$
depends on the frequency $\omega$ and on the orientation of the split-ring  (angle $\theta$).
To analyze this dependance assume that the incident field is parallel to the $y$ axis.
$$\mathbf{f}_{\rm in} =  \left ( \begin{matrix} 0 \cr 1 \cr 0 \end{matrix}
\right ) . $$
Then the reflected and transmitted fields can be written as
$$\mathbf{f}_{\rm r} =\left ( \begin{matrix} 0 \cr R_1 \cr R_2 \end{matrix}
\right ),~~
\mathbf{f}_{\rm tr}= \left ( \begin{matrix} 0 \cr T_1 \cr T_2 \end{matrix}
\right ) . $$
From the relations (\ref{def_l}),(\ref{def_d}),(\ref{fr}),(\ref{ftr}) we get
\begin{eqnarray}
\label{r12} R_1 = T_1-1,~~ R_2 = T_2,\\
T_1 = D(\omega) \left[  1 -i {\kappa \omega \over 2}\sin^2 \theta
\mathcal{M}(\omega) \right ] ,
\nonumber \\
T_2 = - {i \kappa \omega  \sin \theta \cos \theta \over 2} D(\omega)
\mathcal{M}(\omega) . \label{t12}
\end{eqnarray}

\begin{figure}[H]
\centerline{ \resizebox{8 cm}{5 cm} {\includegraphics{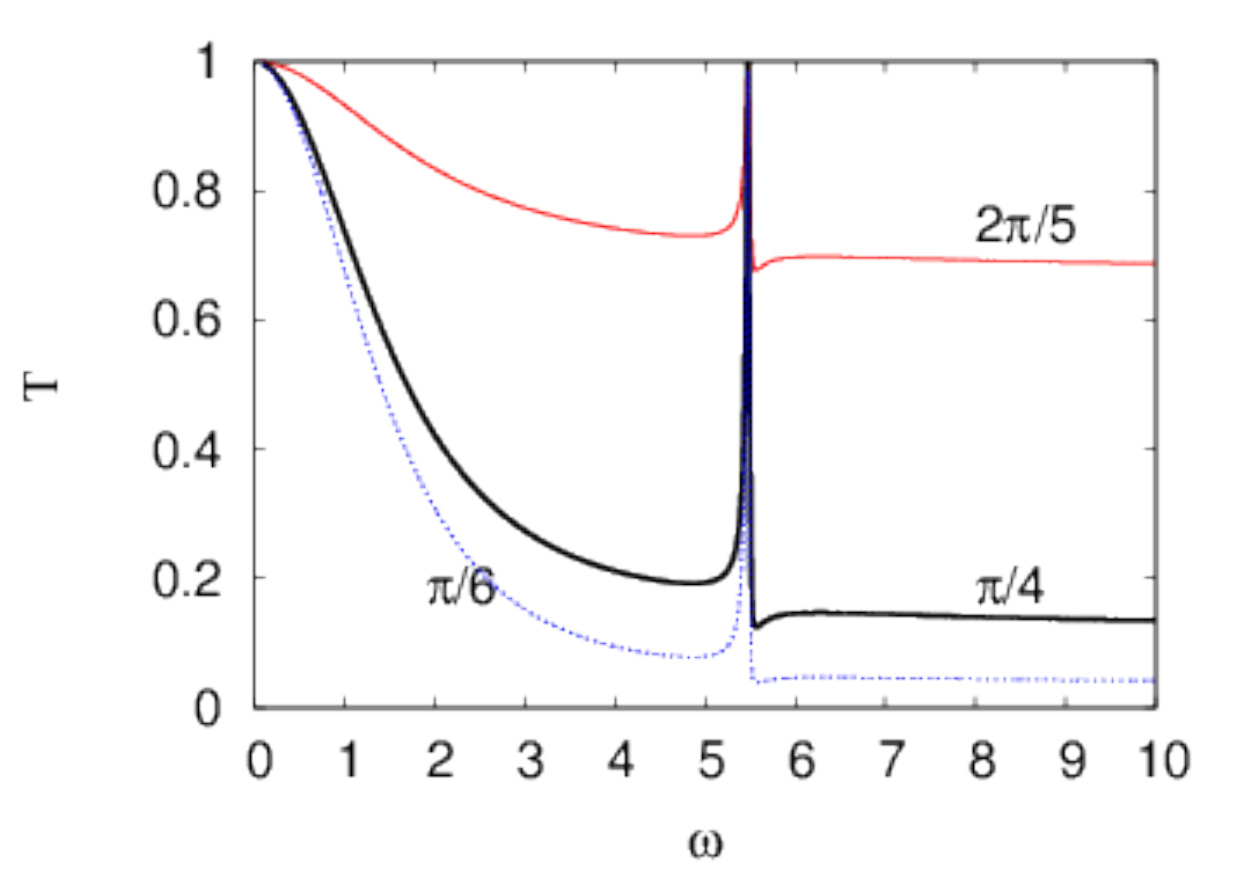} }}
\caption{The transmission coefficient $\mathcal{T}(\omega)$ from
(\ref{trans}) for a plane wave and three values of the
orientation angle $\theta$ .}
\label{transm}
\end{figure}

The transmission coefficient $\mathcal{T}(\omega)$ is shown in Fig.
\ref{transm} for three different values of the orientation angle
$\theta = 2\pi/5, ~~\pi/4$ and $\pi/6$.
Note that the array is transparent $\mathcal{T}=1$ when $\omega=\omega_r$.
Past this frequency, the transmission coefficient 
asymptotically tends to a constant.
For a large angle $\theta = 2\pi/5$, the ring has less influence, since
most of the field is transmitted. When the angle is reduced, the
incoming magnetic field interacts strongly with the rf-SQUIDs so
the reflected field is stronger.

\begin{figure}[H]
\centerline{ \resizebox{8 cm}{5 cm} {\includegraphics{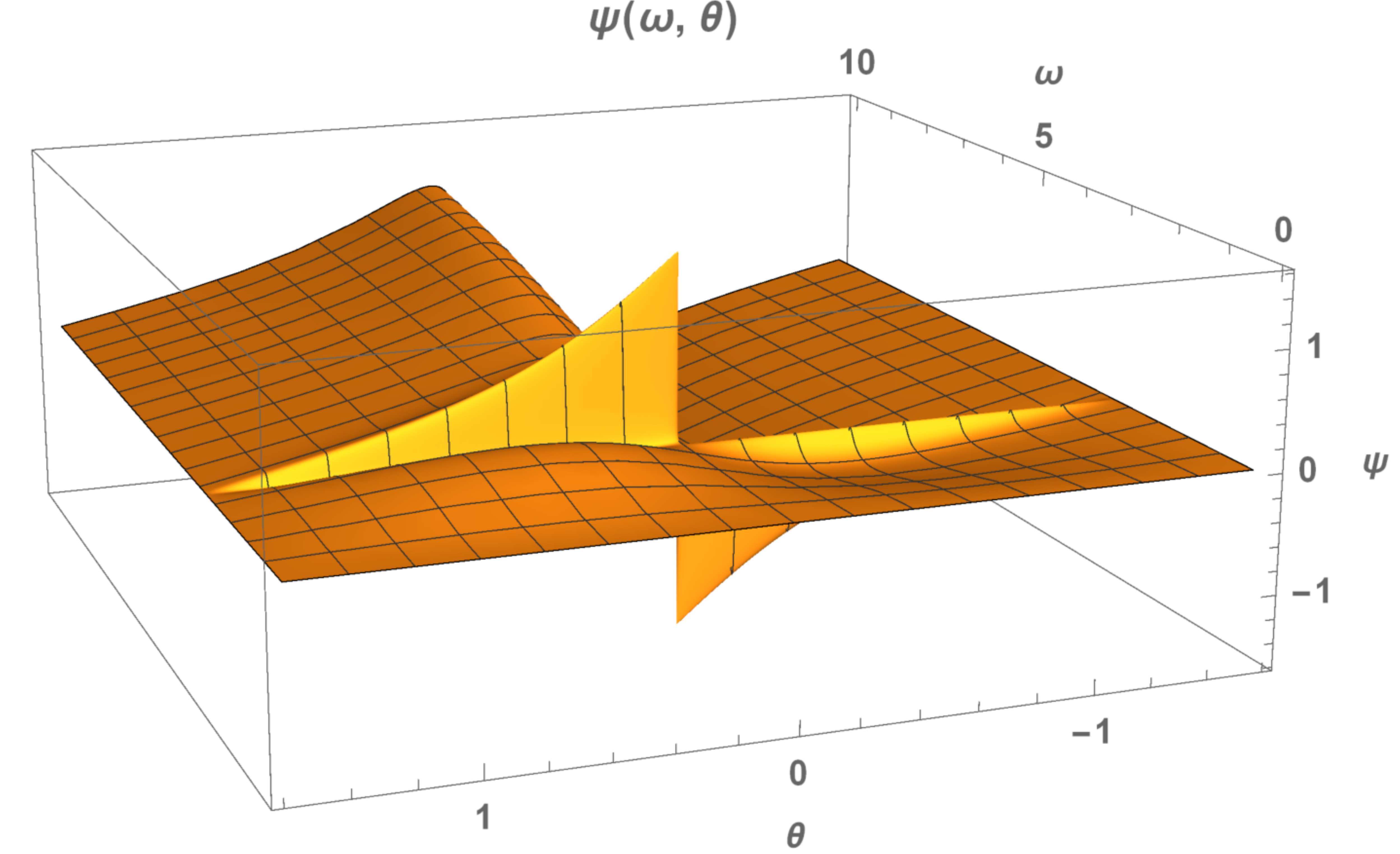} }}
\caption{Angle $\psi (\omega)$ from (\ref{psi}) as function of $\omega$ and $\theta$, $\alpha =0$}
\label{angle}
\end{figure}

From the expression~(\ref{t12}) for $T_{1,2}$ one obtains
the angle of polarization rotation of the transmitted wave $ \psi $
in the $(y,z)$ plane:  
\begin{equation}\label{psi}
    \psi = \arctan \left( \frac{\mathrm{Re}~T_2}{\mathrm{Re}~T_1}\right).
\end{equation}
Fig. \ref{angle} shows this angle $\psi$  as a function of
$\omega$ and $\theta$.
\begin{figure}[H]
\centerline{ \resizebox{8 cm}{5 cm} {\includegraphics{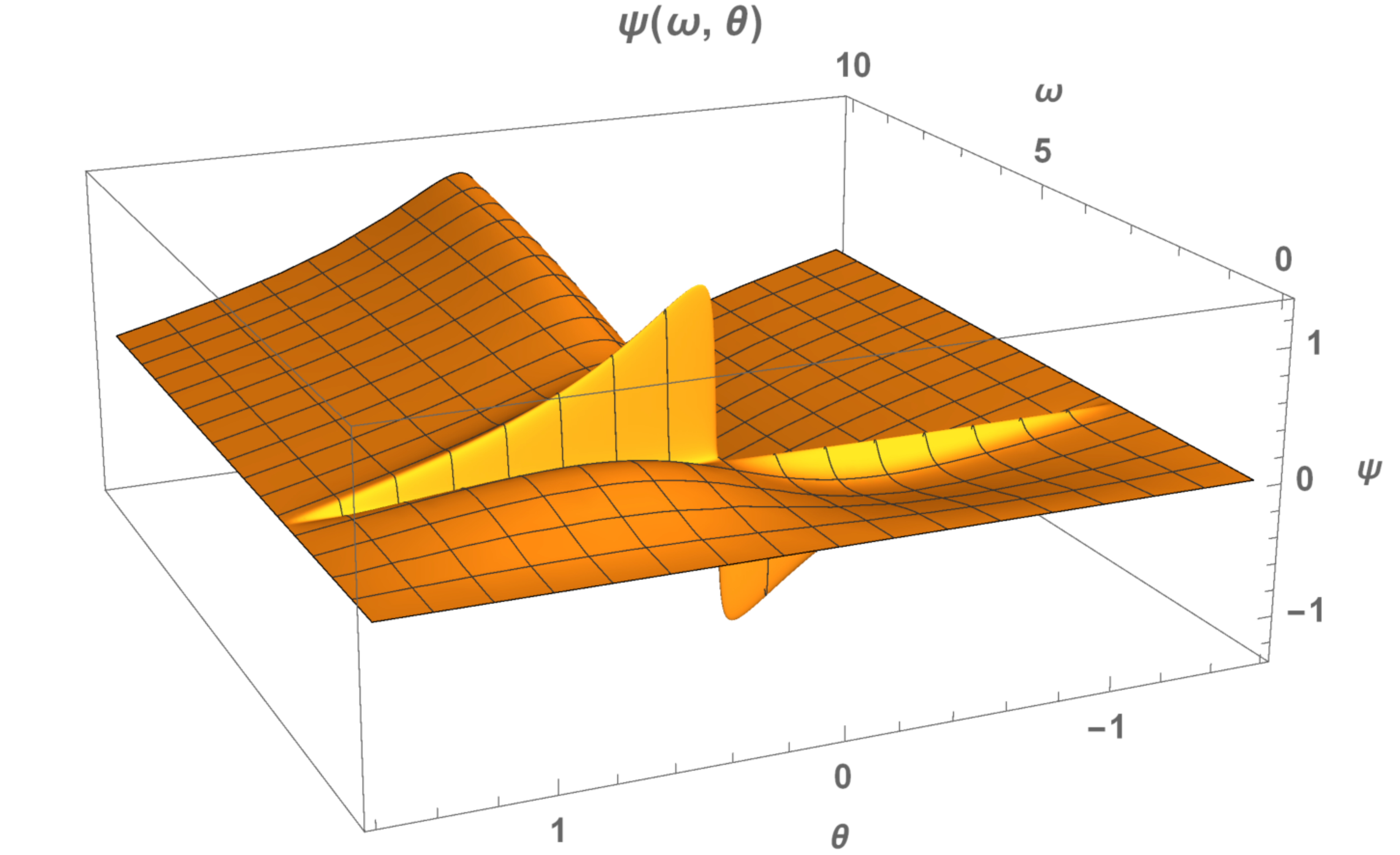} }}
\caption{Angle $\psi (\omega)$ from (\ref{psi}) as function of $\omega$ and $\theta$, $\alpha =0.05$}
\label{angleLoss}
\end{figure}
Fig.~\ref{angle} demonstrates the Faraday effect which takes place in  rf-SQUID metasurface under consideration. The polarization rotation angle $\psi$ depends both on the frequency  $\omega$  and on the  inclination angle of the 
rf-SQUIDs. The angle of rotation of the polarization changes sharply
near the resonant frequency $\omega_r$ (gigantic Faraday effect). 
This angle depends strongly on $\theta$; it has a pronounced resonance 
character near $\omega_r$.
In this case, when $\theta$ changes from $-\pi/2$ to $0$ angle $\psi$ monotonically decrease from $0$ to $-\pi/2$. For $\theta =0$ the angle $\psi$ 
jumps from $-\pi/2$ to $\pi/2$ and monotonically decreases to $0$ when 
$\theta$ changes from $0$ to $\pi/2$ (see Fig. ~\ref{angle}). Introducing
losses which are always present in any practical situation smoothes the
jump transition, Fig.~\ref{angleLoss}.

\section{Detailed study of the resonance}

We study here the characteristics of the resonance of the
coupled system -- rf-SQUID and field.  From equations
(\ref{djj}) and (\ref{ftr}) we can write the time dependent phase equation as~(\cite{Maim:Gabi:10})
\begin{eqnarray}
&~& \ddot{\varphi}(t) +\alpha \dot{\varphi}(t) + (1+\beta \cos\phi_s) \varphi (t)=-f(t),\nonumber\\
&~& f(t)={\kappa \over 2 } \left(\dot{\varphi}(t)+\dot{f}(t) \right) + k\cos \omega t. \label{init:syst}
\end{eqnarray}
Recalling the resonant frequency $\omega_{r}$ from (\ref{reson}) we replace
$$\cos \omega t \mapsto \exp (i\omega t).$$
so that the system~(\ref{init:syst}) can be rewritten:
The solutions of the resulting system have the following form:
\begin{eqnarray}
&\varphi(t)=A\exp(i\omega t),~~ f(t)=B \exp(i\omega t) , \nonumber\\
& A=-kC^{-1}, ~~ B=k\left[(\omega_{r}^2 -\omega^2)+i  \alpha \omega \right]C^{-1} , \nonumber\\
& C = \omega_{r}^2 -\omega^2 (1-{\kappa \over 2 }\alpha) +i\omega
\left[\alpha+{\kappa \over 2 }(1+\omega^2 - \omega_{r}^2) \right]
\nonumber.
\end{eqnarray}
Representing $A$ in terms of a phase and an amplitude $A=a\exp(i \theta)$
we obtain:
\begin{eqnarray}
& a=\frac{k}{\sqrt{\left[\omega_{r}^2 -\omega^2 (1-(\kappa\alpha/2))\right]^2
+\omega^2  \left[\alpha+(\kappa/2)(1+\omega^2 - \omega_{r}^2) \right]^2}} , \nonumber\\
& \tan \theta =\frac{\omega \left[\alpha +(\kappa/2)(1+\omega^2
-\omega_{r}^2)\right]}{\omega^2 (1-(\kappa\alpha/2)-\omega_{r}^2}
.\nonumber \label{complex2}
\end{eqnarray}
\begin{figure}[H]
\centerline{ \resizebox{10 cm}{5 cm} {\includegraphics{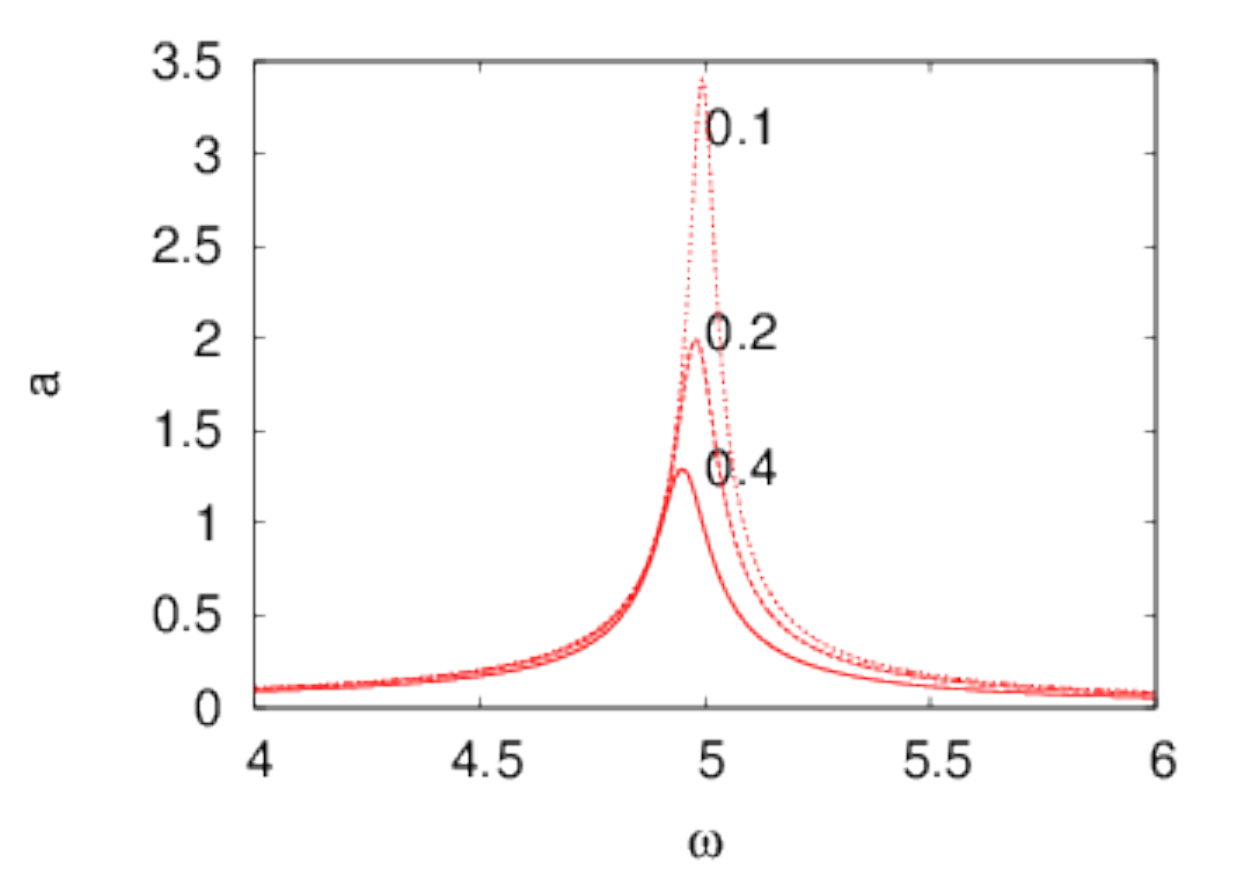} \includegraphics{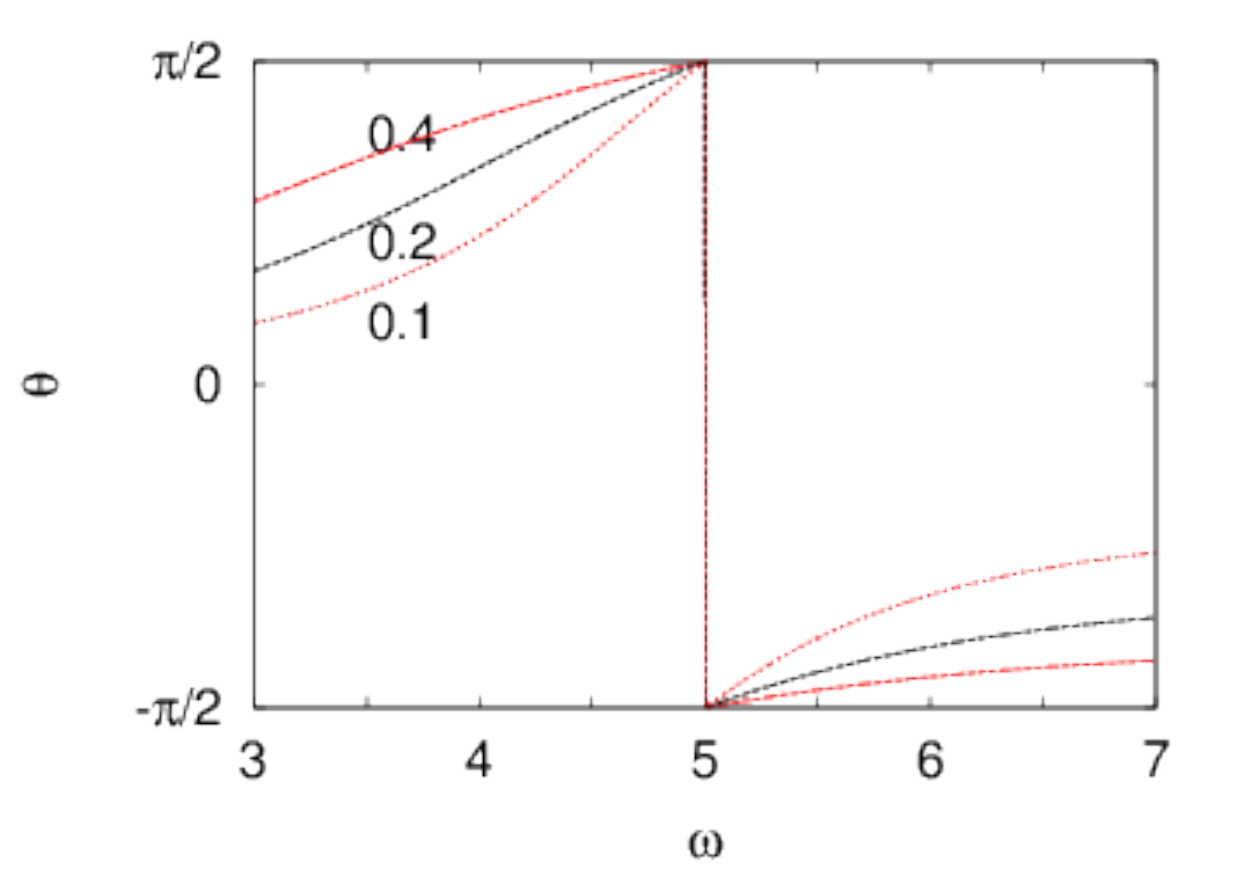}}}
\caption{ Plot of the amplitude $a(\omega)$ (left panel) and the phase
$\theta(\omega)$ (right panel) for three different values of the
coupling $\kappa =0.4 , 0.2 $ and $0.1$. The other parameters
are $\omega_r=5, \alpha=0.01$.}
\label{tanta}
\end{figure}
The amplitude $a$ and the phase $\theta$ are plotted as a function
of the frequency $\omega$. Notice how the resonance gets sharper for
a small coupling $\kappa$. We also have the typical jump in the phase
as one crosses the resonance.

The amplitude $a(\omega)$ reaches its maximal value for
\begin{eqnarray}
&&\omega_{max}^{2}=
\frac{1}{3{\tilde{\kappa} }^2}\left[ {\tilde{\kappa} }^2 (2 \omega_{r}^2 -\alpha^2 -2 ) -1\right]+ \nonumber\\
&& +\frac{1}{3{\tilde{\kappa} }^2}
\bigg[1 +{\tilde{\kappa} }^2 \left[ 4+ 2\omega^{2}_{r} - \alpha^2 -6{\tilde{\kappa} }\alpha +\right.\nonumber \\
&& \left. {\tilde{\kappa} }^2 \left[( \omega^{2}_{r}-1)\left[(\omega^{2}_{r}-1)-4\alpha^2\right] +\alpha^4 \right] \right] \bigg]^{1/2}
\end{eqnarray}
where $\tilde{\kappa}=\kappa /2 $.  Since $\alpha$ and
${\tilde{\kappa} }$ are small, $\omega_{max}^{2}$ up to second order
has following form:
$$
  \omega_{max}^{2}\approx \omega_{r}^2 -\frac{\alpha^2}{2} -\alpha {\tilde{\kappa} }-\left(\omega_{r}^2+\frac{1}{2}\right){\tilde{\kappa} }^2
$$
When $\omega$ is close to $\omega_r$, ($\omega =\omega_r +\Delta $,
$|\Delta |\ll 0$) and $\alpha$ and ${\kappa }$ are small, then
\begin{equation}\label{complex4}
    a\approx \frac{k}{\omega_{r}\sqrt{4 \Delta^{2}
+(\alpha +{\tilde{\kappa} })^2 }}
\end{equation}

If ${\kappa }=0$, we recover the classical resonance observed for a
damped driven linear oscillator, both for $\theta$ and $a$,
\cite{landau}.

\section{Conclusion}

We analyzed the interaction of a plane
electromagnetic wave with a two dimensional array of rf-SQUIDs. The
wave vector of the incident field is assumed orthogonal
to the array of rf-SQUIDs. All rf-SQUIDs have the
same inclination with respect to the surface of the array and
the effective thickness of this "array film" is much smaller then the
wavelength.
Our main result is that despite this small thickness, the array
effectively controls the wave reflection and transmission.
In particular, it changes the polarization of the reflected
wave and this change is
determined only by the orientation of the rf-SQUIDs. This effect is
identical to the Kerr effect in a gyrotropic medium. Here the 
gyrotropy is introduced by the rf-SQUIDs.  The polarization of the
transmitted wave also changes and depends both on the carrier
frequency and on the inclination angle of the rf-SQUIDs. This
is similar to a Faraday effect. At the resonance frequency we obtain
a gigantic Faraday effect. 
We emphasize that the thickness of the array is always much
smaller than the wave-length.

This array of rf-SQUIDs acts as a meta-surface
that controls the polarization of an electromagnetic wave.
The analysis that we carried out is
limited by the linear approximation. Increasing the incident field
will cause a larger current in the ring and subsequently
a nonlinear response of the rf-SQUIDs. We then expect
nonlinear Kerr and Faraday effects combined with bistability.

\section{Acknowledgement}

The research of A.I.M. was supported by Russian Scientists Found
(project 14-22-00098). The research of I.R.G. was partially
supported by the Ministry of Education and Science of Russian
Federation (Project DOI: RFMEFI58114X0006). J. G. C. thanks the
Region Haute-Normandie for support through a research grant GRR-LMN
and the CRIHAN computing center for the use of its facilities.


\begin{thebibliography}{99}

\bibitem{ygk11}
\textit{N. Yu et al}, "Light propagation with phase discontinuities:
generalized laws of reflection and refraction", Science,
\textbf{334}, 333-337, (2011).

\bibitem{shalaev12}
\textit{X. Ni et al }, "Broadband light bending with plasmonic
nanoantennas", Science, \textbf{335}, 427, (2012).

\bibitem{Lazarides:07a}
\textit{N. Lazarides, G. P. Tsironis, and M. Eleftheriou},
"Dissipative discrete breathers in rf SQUID metamaterials",
arXiv:0712.0719v1 [nlin.PS] 2007

\bibitem{Lazarides:07b}
\textit{N. Lazarides, and G. P. Tsironis}, "rf superconducting
quantum interference device metamaterials", Appl. Phys. Lett.
\textbf{90}, 163501 (2007).

\bibitem{Lazarides:09} \textit{G. P. Tsironis, N. Lazarides, and M. Eleftheriou},
"Dissipative Breathers in rf SQUID Matamaterials" PIERS Proc. March
23-27, China, Beijin 2009 p. 52-56 (2009).

\bibitem{Maim:Gabi:10} \textit{A.I. Maimistov, I. Gabitov},
Nonlinear response of a thin metamaterial film containing Josephson
junctions, Opt.Commun. \textbf{283}, 1633-1639 (2010).



\bibitem{Chunguang:06} \textit{Chunguang Du, Hongyi Chen, and Shiqun Li}, "Quantum left-handed
metamaterial from superconducting quantum-interference devices",
Phys.Rev. \textbf{B74}, 113105 (2006).

\bibitem{Anlage:11} \textit{St.M Anlage}, "The physics and applications of superconducting
metamaterials", J.Opt. \textbf{13}, 024001 (2011).

\bibitem{Trepanier:11} \textit{M. Trepanier, Daimeng Zhang, O.
Mukhanov, and St. M. Anlage}, "Realization and Modeling of
Metamaterials Made of rf Superconducting Quantum-Interference
Devices",  Phys.Rev. \textbf{X 3}, 041029 (2013).

\bibitem{cgm12}
\textit{J.-G.~Caputo, I. Gabitov  and A.I.~Maimistov },
"Electrodynamics of a split-ring Josephson resonator in a microwave
line", Phys Rev. \textbf{B 85}, 205446  (2012).


\bibitem{MFK:10}\textit{A.E. Miroshnichenko, S. Flach, Y.S. Kivshar}, Fano resonances in nanoscale structures, Reviews of Modern Physics \textbf{82} (3), 2257, (2010)

\bibitem{Ustinov:13} \textit{P. Jung, S. Butz, S. V. Shitov, and A. V. Ustinov},
"Low-loss tunable metamaterials using superconducting circuits with
Josephson junctions", Appl.Phys.Lett. \textbf{102}, 62601 (2013).







\bibitem{Lazarides:08} \textit{N. Lazarides, G.P. Tsironis, and Yu.S. Kivshar}
"Surface breathers in discrete magnetic metamaterials", Phys.Rev.
\textbf{E 77}, 065601(R) (4 pages) (2008).

\bibitem{Elefth:08} \textit{M. Eleftheriou, N. Lazarides,  G.P.Tsironis} "Magnetoinductive
breathers in metamaterials", Phys.Rev. \textbf{E77}, 036608 (13
pages) (2008).

\bibitem{Elefth:Lazar:09} \textit{M. Eleftheriou, N. Lazarides,
G.P.Tsironis, and u.S. Kivshar} "Surface magnetoinductive breathers
in two-dimensional magnetic metamaterials", Phys.Rev. \textbf{E80},
 017601 (4 pages) (2009).

\bibitem{landau} L. Landau and E. Lifchitz, "Cours de physique th\'eorique",
Mecanique Mir, Moscou.


\end{thebibliography}
\end{document}